\documentclass[aps,
prb,
reprint,
preprintnumbers,
superscriptaddress,
amsfonts,
amssymb,
amsmath
]{revtex4-2}
\usepackage{bm,latexsym,mathrsfs,enumerate}
\usepackage{upgreek}
\usepackage[table,x11names]{xcolor}
\usepackage[breaklinks=true,
unicode=true,
urlcolor = RoyalBlue4,
colorlinks = true,
citecolor = RoyalBlue4,
linkcolor = blue
]{hyperref}
\usepackage{graphicx}
\usepackage{lmodern}
\usepackage{mathtools}
\usepackage{pifont}
\usepackage{multirow}
\usepackage[cal=boondoxo,frak=boondox,scr=rsfs]{mathalfa}
\usepackage{savesym}
\usepackage[most]{tcolorbox}
\savesymbol{comment}
\usepackage[todonotes={textsize=tiny,textwidth=45pt},truncate=hyphenate,deletedmarkup=xout,addedmarkup=uwave,commandnameprefix=ifneeded]{changes}
\renewcommand{\vec}[1]{\bm{#1}}
\DeclareMathOperator{\sign}{sign}
\usepackage[final]{pdfpages}
\usepackage{tikz}
\makeatletter
\AtBeginDocument{\let\LS@rot\@undefined}
\makeatother
\begin{document}
\title{Control of magnetic response in curved stripes by tailoring cross-section}

\author{Kostiantyn V. Yershov}
\email{k.yershov@ifw-dresden.de}
\affiliation{Leibniz-Institut f\"ur Festk\"orper- und Werkstoffforschung, IFW Dresden, 01171 Dresden, Germany}
\affiliation{Bogolyubov Institute for Theoretical Physics of the National Academy of Sciences of Ukraine, 03143 Kyiv, Ukraine}

\author{Denis D. Sheka}
\email{sheka@knu.ua}
\affiliation{Taras Shevchenko National University of Kyiv, 01601 Kyiv, Ukraine}

\date{January 6, 2023}




%
%

\begin{abstract}

Curved magnetic architectures are key enablers of the prospective magnetic devices with respect to size, functionality and speed. By exploring geometry-governed magnetic interactions, curvilinear magnetism offers a number of intriguing effects in curved magnetic wires and curved magnetic films. The applicability of the current micromagnetic theory requires that the sample has constant width and thickness, which does not correspond in many cases to specificity of experimental sample preparation. Here, we put forth a self-consistent micromagnetic framework of curvilinear magnetism of nanowires and narrow stripes with spatially inhomogeneous cross-section. The influence of the varying cross-section is exploited and illustrated by an example of the simplest topological texture, which is a transversal head-to-head (tail-to-tail) domain wall. The cross-section gradient becomes a source of domain wall pinning which competes the curvature gradient. Eigenfrequencies of the domain wall free oscillations at the pinning potential are determined by both curvature and cross-section gradients. Prospects for curvilinear magnonics and spintronics are discussed.
\end{abstract}

\maketitle


\emph{Introduction}. Manipulation of the material response of objects using its geometrical properties became an important topic of contemporary physics. In case of magnetism a mutual interplay of magnetization texture (material properties), curvature and topology (geometrical properties) becomes a playground of curvilinear magnetism \cite{Streubel16a}. This rapidly developing research area of modern magnetism is aimed to explore geometry-induced effects in curved magnetic wires and films. Active exploration of this new material class turns light on the fundamentals of magnetism of nanoobjects with curved geometry and applications of 3D-shaped curved magnetic nanoarchitectures, leading to remarkable developments in shapable magnetoelectronics, magnetic sensorics, spintronics, 3D magnonics, and microrobotics \cite{Makarov22a}. 

An existing micromagnetic framework of curvilinear magnetism requires that samples possess inalterable cross-section with constant width and thickness \cite{Makarov22a, Makarov22, Sheka22}. In real experiments width and thickness of samples can vary in a wide range. In particular, thickness gradients of 2D film are often achieved using moving shutters \cite{Vishwakarma20}, plasma-enhanced chemical vapor deposition methods \cite{Lopez-Santos17}. In the nanosphere lithography there appears a thickness gradient across the cap structure with the thickest film at the top of the cap and thinnest film close to the equator due to the specificity of the sample preparation using magnetron sputtering \cite{Ulbrich06, Ulbrich10}. The specially varying width of the planar stripes in the form of notches and protrusions can pin or even boost domain walls \cite{Himeno05, Petit08, Garcia-Sanchez11, Yuan15}; in asymmetric nanorings changes in the stripe width become geometrical sources of domain wall nucleation \cite{Richter16} and its automotion \cite{Mawass17}. The highly spatially varying thickness gradients achieved with 3D nanopatterning using focused-electron-beam-induced deposition (FEBID). In particular, namely the thickness gradient is expected to be a dominant mechanism of experimentally observed domain wall automotion in 3D interconnectors \cite{Skoric22}. The space modulation of diameter in nanotubes becomes the main source of the domain wall pinning \cite{Fernandez-Roldan19}.

Here, we present generalized micromagnetic framework of curvilinear wires and stripes with varying cross-section, e.g. with thickness and (or) width gradients. This theory allows not only to predict novel geometry-induced effects in conventional materials, but also explain recent experiments \cite{Petit10, Mawass17, Skoric22} and propose applied routes to explore the utility of 3D-shaped curved magnetic architectures for curvilinear spintronics and curvilinear magnonics. We apply theory to predict novel effects in static and linear dynamics of domain walls in curved stripes with varying cross-section. In particular, the domain wall can be pinned by the local cross-section deformation and eigenfrequency of the domain wall free oscillations at the pinning potential are determined by both curvature and cross-section gradients.


\emph{Results}. We aim to construct a theory of thin curved ferromagnet wires and stripes with varying cross-section. For this purpose we represent a ferromagnet body as a space domain $\vec{r} = \vec{\gamma} + \zeta_1 \vec{e}_\textsc{n} + \zeta_2 \vec{e}_\textsc{b}$. Here, $\vec{\gamma} = \vec{\gamma}(s)$ described the central curve with $s$ being the arc length coordinate, $\zeta_1$ and $\zeta_2$ parameterize the perpendicular cross-section with varying area $S=S(s)$, unit vectors $\vec{e}_\textsc{n}$ and $\vec{e}_\textsc{b}$ determine normal and binormal direction to the central curve, respectively. Let us focus on a classical biaxial ferromagnet with the simplest energy density, $W=W^{\textsc{x}} + W^{\textsc{a}}$. Here, $W^{\textsc{x}} = -A\vec{m}\cdot\!\vec{\nabla}^2\vec{m}$ is the exchange energy density with $A$ being the exchange constant, $\vec{m}=\vec{M}/M_{\textsc{s}}$ being the normalized magnetization, and $M_{\textsc{s}}$ being the saturation magnetization. The next term, $W^{\textsc{a}} = - K_{\textsc{t}}^\text{eff} \left(\vec{m}\cdot \vec{e}_{\textsc{t}}\right)^2 + K_{\textsc{b}}^\text{eff}\left(\vec{m}\cdot \vec{e}_{\textsc{b}}\right)^2$ described the density of biaxial anisotropy energy with $K_\textsc{t}^\text{eff} = K_\textsc{t}+4\pi M_s^2 k_\textsc{t}^\text{ms}$ and $K_\textsc{b}^\text{eff} = K_\textsc{b}+4\pi M_s^2 k_\textsc{b}^\text{ms}$ being anisotropy coefficients of tangential easy axis and binormal hard axis, respectively, $K_\textsc{t} > 0$ and $K_\textsc{b} > 0$ being magnetocrystalline anisotropy constants. Terms $k_\textsc{t}^\text{ms}$ and $k_\textsc{b}^\text{ms}$ arise from the magnetostatic contribution, see 
Supplementary Section SM-1~\footnote[1]{See Supplemental Material \texttt{link provided by the publisher} for details of the analytical calculation, micromagnetic simulations, and movies, which includes Refs.~\cite{Carmo16,Kuhnel05,Hillebrands06,Porter04,Malozemoff79,Slonczewski72,Aharoni98,Gaididei17a,Mougin07,Yershov18a,Kravchuk14c,Slastikov12,Gaididei14,Sheka15,Sheka20a,Sheka21b,Yershov15b,Fischbacher07}.} for details.

The energy of the curved magnet is well-known \cite{Makarov22a} to be restructured in the curvilinear reference frame, which follows the sample geometry, providing means to recover the translation invariance of the effective anisotropy. The total energy, normalized by the $E_0=4\pi M_{\textsc{s}}^2\ell^3$, has the form $\mathcal{E}=E/E_0= \int \mathscr{E}(\xi) \mathrm{d}\xi$ with the energy density
\begin{equation} \label{eq:energy}
\mathscr{E}=\mathscr{E}_0 + \underbrace{\mathscr{E}_{\textsc{a}}}_{\substack{\text{effective}\\ \text{anisotropy}}} + 
\underbrace{\mathscr{E}_{\textsc{d}}}_{\substack{\text{effective}\\ \text{DMI}}}\hspace{-0.5em}.
\end{equation}
Here an exchange length $\ell=\sqrt{A/\left(4\pi M_{\textsc{s}}^2\right)}$ determines a length scale of the system.  The applicability of the current micromagnetic framework requires that anisotropy directions do not vary along the cross-section, i.e. $\vec{e}_\textsc{t} = \vec{e}_\textsc{t}(s)$ and $\vec{e}_\textsc{b} = \vec{e}_\textsc{b}(s)$. This makes it possible to suppose that the magnetization texture remains uniform along the direction of the normal to the sample, which typically means that the sample thickness does not exceed several times the characteristic magnetic length scale $\ell$. One more restriction is that possible deformations of the sample cross-section are smooth enough~\cite{Petit10,Mawass17,Skoric22}. These limitations specify the applicability of the theory to the description of quasi-1D objects including curved wires, stripes, and ribbons.

The first energy contribution in \eqref{eq:energy}, the term $\mathscr{E}_0 = \mathcal{S} m_i'm_i'$, is a `common', regular isotropic part of exchange interaction with $\mathcal{S}=S(\xi)/\ell^2$ being dimensionless cross-section and $\xi=s/\ell$ being the dimensionless coordinate along the central curve of the sample. The Einstein summation convention is applied here and below, prime denotes the derivative with respect to $\xi$, and indices $i$, $j$, $k$ numerate curvilinear coordinates and curvilinear components of magnetization. The second term, an effective anisotropy, $\mathscr{E}_{\textsc{a}} = \mathcal{K}_{ij} m_i m_j$ comprises the intrinsic magnetocrystalline anisotropy $W^{\textsc{a}}$ and extrinsic curvilinear geometry-governed contributions. Effective anisotropy coefficients are $\mathcal{K}_{ij} = \mathcal{S} \left(\vec{\varpi}^2\delta_{ij} -\varpi_i\varpi_j - k_\textsc{t}\delta_{1i}\delta_{1j} + k_\textsc{b}\delta_{3i}\delta_{3j}\right)$ with $\vec{\varpi} = \sigma\vec{e}_\textsc{t} +\varkappa\vec{e}_\textsc{b}$ being the Darboux vector, determined by reduced curvature $\varkappa = \kappa\ell$ and reduced torsion $\sigma=\tau \ell$, and $\delta_{ij}$ being Kronecker delta; the reduced anisotropy coefficients $k_\textsc{t}$ and $k_\textsc{b}$ are determined in Supplementary Section SM-1~\cite{Note1}. An effective geometry-governed Dzyaloshinskii–Moriya interaction (DMI) $\mathscr{E}_{\textsc{d}} = \varepsilon_{ijk} \mathcal{D}_i m_j m_k'$ is linear with respect to curvature and torsion with $\vec{\mathcal{D}} = 2\mathcal{S} \vec{\varpi}$ being geometry-governed exchange-driven Dzyaloshinskii vector.


\begin{figure}
\includegraphics[width=\columnwidth]{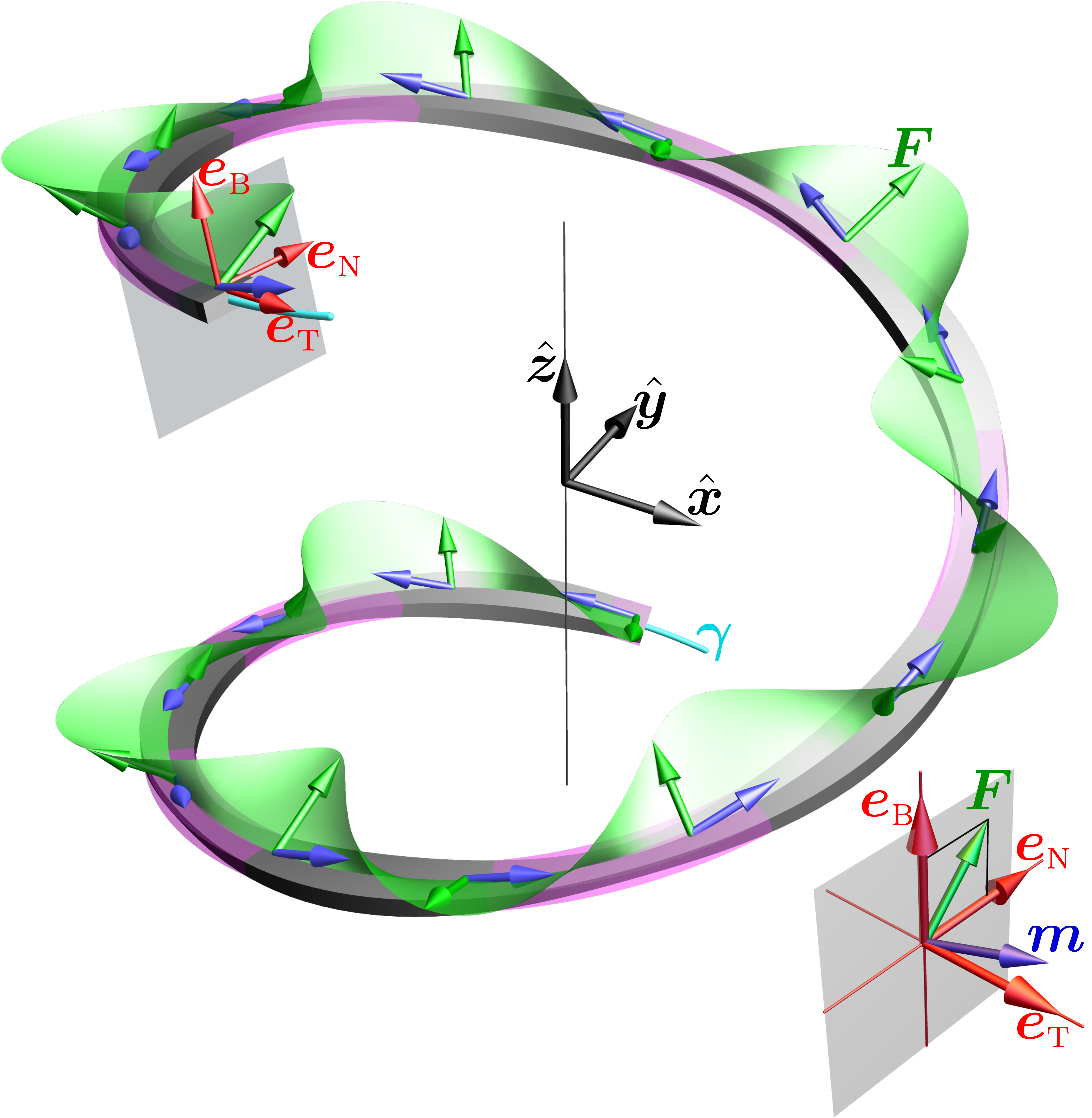}
\caption{(Color online) \textbf{Emergent geometry-induced magnetic field}:	
Schematics of emergent field $\vec{F}$ and equilibrium magnetization texture $\vec{m}$ in a curved stripe $\vec{r}$ with spatially varying cross-section. Transparent magenta color corresponds to the stripe shape without cross-section deformation. Green arrows and ribbon correspond to the direction of the field $\vec{F}$, blue arrows correspond the equilibrium magnetization distribution, and cyan line corresponds to the central line $\vec{\gamma}$ of the stripe.
\label{fig:emergent_field}%
}
\end{figure}

\emph{Emergent geometry-induced magnetic field}. Let us discuss here the behavior of systems with a strong anisotropy. This allows us to assume that the magnetic texture, modified by the geometry, will not deviate significantly from the equilibrium state given by the anisotropy. By introducing small deviations from strictly tangential magnetization distribution, one can obtain energy density~\eqref{eq:energy} in the following form, see Supplementary Section SM-2~\cite{Note1} for details:
\begin{equation} \label{eq:emergent-field}
\begin{aligned}
\mathscr{E}& \approx \mathscr{E}_\textsc{t} - \mathcal{S}\vec{F}\cdot\vec{m} + \mathcal{S}\left[k_\textsc{t}\left(\vartheta^2+\varphi^2\right)+k_\textsc{b}\vartheta^2\right],\\
\vec{F} &= 2\mathfrak{C} \left(\varkappa'+\varkappa \frac{\mathcal{S}'}{\mathcal{S}} \right) \vec{e}_\textsc{n} + 2\mathfrak{C}\varkappa\sigma\,\vec{e}_\textsc{b}.
\end{aligned}
\end{equation}
Here, the first term $\mathscr{E}_\textsc{t}$ is the energy density of the strictly tangential distribution, the second term describes the interaction with \textit{emergent geometry-induced} magnetic field~$\vec{F}$, see Fig.~\ref{fig:emergent_field}. Parameter $\mathfrak{C}=\pm 1$ defines the direction of magnetization along the stripe: $\mathfrak{C}=+1$ corresponds to parallel orientation along $\vec{e}_\textsc{t}$ and $\mathfrak{C}=-1$ corresponds to the antiparallel. This field causes the  magnetization tilting from the tangential distribution by the angles
\begin{equation} \label{eq:tilting}
\vartheta\approx-\frac{\mathfrak{C}}{k_\textsc{t}+k_\textsc{b}}\varkappa\sigma,\quad \varphi\approx \frac{1}{k_\textsc{t}} \left(\varkappa'+\varkappa\frac{\mathcal{S}'}{\mathcal{S}}\right).
\end{equation}
One can see that the cross-section gradient $\mathcal{S}'$ acts as a \emph{geometrical source} of the ground state tilting in addition to the curvature gradient \cite{Gaididei14, Sheka15, Sheka22}. While the strictly tangential magnetization distribution in curved wires/stripes with the constant cross-section can be realized only in straight or flat arc-shaped wires with constant curvature, samples with varying cross-section possess another criterion: the strictly tangential distribution is possible when $\varkappa'/\varkappa=-\mathcal{S}'/\mathcal{S}$ and $\sigma=0$ only.


\textit{Domain wall in a planar curved stripe. -- }We illustrate the above theory by a flat narrow curved ferromagnetic stripe of a rectangular cross-section. Using curvilinear reference frame, one can parameterize the magnetization as $\vec{m}=\vec{e}_\textsc{t}\cos\theta+\vec{e}_\textsc{n}\sin\theta\cos\phi+\vec{e}_\textsc{b}\sin\theta\sin\phi$. The spatial-temporal evolution of magnetization follows well-known Landau--Lifshitz--Gilbert equations, its curvilinear form is represented in Supplementary Section SM-3~\cite{Note1}. 

We start with static case, when the minimization of the energy results in a planar texture within the stripe plane with $\cos\phi =\mathcal{C}=\pm1$ and planar deviations from the tangential direction described by $\theta(\xi)$, which satisfies the driven dissipative nonlinear pendulum equation
\begin{equation} \label{eq:theta-def}
\!\theta''+\frac{\mathcal{S}'}{\mathcal{S}}\theta' - k_\textsc{t}\cos\theta\sin\theta\!=\! f(\xi),\;
f(\xi) \!=\! -\mathcal{C}\!\! \left(\!\!\varkappa' + \varkappa \frac{\mathcal{S}'}{\mathcal{S}}\!\right)\!\!.\!\!
\end{equation}
The spatially dependent external force $f(\xi)$ results in the absence of strictly tangential magnetization pattern. This force  has two sources: gradient of the curvature and the gradient of the cross-section, cf.~\eqref{eq:tilting}. Besides, variable thickness causes an effective dissipative motion of the nonlinear pendulum: the term with the first derivative $\theta'$. Note that one can avoid the appearance of effective dissipation by reducing to the parametric pendulum problem, for details see Supplementary Section SM-4 \cite{Note1}.

Let us analyze how this force influences the nonlinear magnetization texture, the domain wall. We apply a collective variable approach based on the $q-\Phi$ model~\cite{Malozemoff79,Slonczewski72}
\begin{equation} \label{eq:qPhi-model}
\cos\theta^{\textsc{dw}}=-p\tanh\left[\frac{\xi-q(t)}{\Delta}\right],\quad \phi^{\textsc{dw}}=\Phi(t).
\end{equation}
When the force is absent, $f(\xi)=0$, this model provides an exact solution of \eqref{eq:theta-def} for the straight stripe with constant cross--section; it describes head-to-head or tail-to-tail domain walls with domain wall width $\Delta=1/\sqrt{k_\textsc{t}}$ for $p = 1$ and $p = -1$, respectively.

\begin{figure}
\includegraphics[width=\columnwidth]{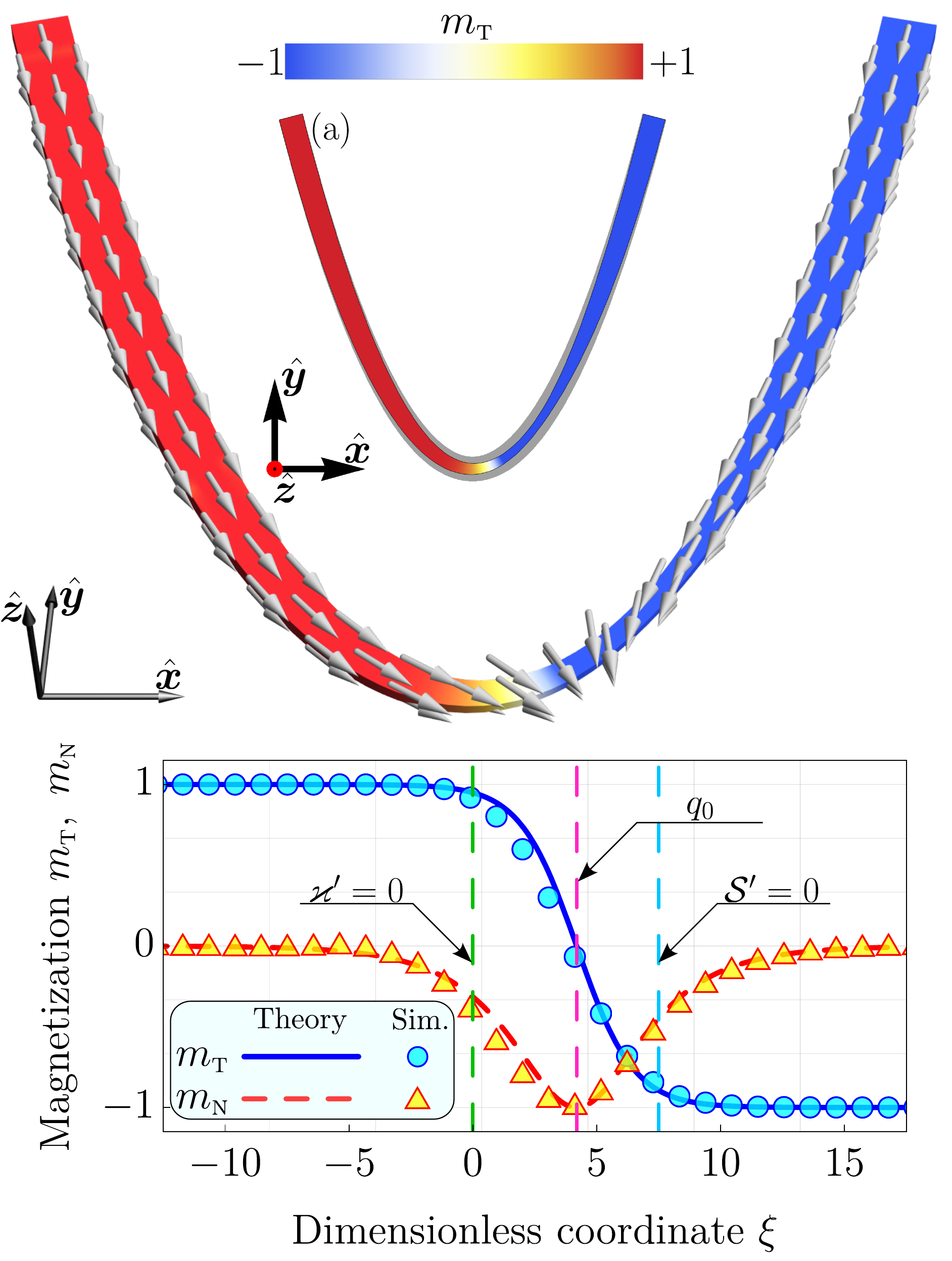}
\caption{(Color online) \textbf{Domain wall pinning}:	
Equilibrium state of the transversal domain wall in a parabola shaped stripe with curvature at the extremum $\varkappa_0=0.2$. The cross-section deformation is defined as $\mathcal{S}/\mathcal{s}_0 = 1 - \varrho/\cosh\left[\left(\xi - \eta\right)/\lambda\right]$ with $\varrho = 0.9$, $\eta = 7.5$, and $\lambda = 15$. Top inset corresponds to the magnetization distribution obtained by means of Nmag micromagnetic simulations, while bottom inset corresponds to comparison of simulations (symbols) and analytical predictions (solid and dashed lines). Vertical magenta dashed line is obtained with prediction~(S15). Inset (a) shows the top view of the stripe geometry where gray color corresponds to the stripe with constant cross section.
\label{fig:equilibrium}%
}
\end{figure}

The domain wall motion can be realized under the influence of the force $f(\xi)$. This dynamics can be described using collective variables $\{q,\Phi\}$, which determine the domain wall position and phase, respectively. The domain wall width $\Delta$ is assumed to be a slaved variable~\cite{Hillebrands06} i.e., $\Delta=\Delta\left[q(t),\Phi(t)\right]$. Such an approach is valid, when the force $f(\xi)$ can be considered as a small perturbation, which does not modify significantly the domain wall profile, i.e. when the curvature gradients and the cross-section gradients are weak on a scale of domain wall width. In equilibrium narrow domain wall becomes pinned at the position $q_0$ and its angle $\Phi_0$, which are determined by:
\begin{subequations} \label{eq:equilibrium}
\begin{align} \label{eq:q0}
&\frac{\varkappa'_0}{\varkappa_0}+\frac{\mathcal{S}'_0}{\mathcal{S}_0}=\frac{\sqrt{k_{\textsc{t},0}}}{\pi|\varkappa_0|} \left(\frac{k_{\textsc{t},0}'}{k_{\textsc{t},0}}+2\frac{\mathcal{S}'_0}{\mathcal{S}_0}\right),\\
\label{eq:equilibrium-Phi0} %
&\cos\Phi_0 =-p \sign \varkappa_0, \quad \Delta_0 = 1/\sqrt{k_{\textsc{t},0}},
\end{align}
\end{subequations}
where $\varkappa_0\equiv \varkappa(q_0)$, $\mathcal{S}_0\equiv \mathcal{S}(q_0)$, and $k_{\textsc{t},0}\equiv k_{\textsc{t}}(q_0)$, see Supplementary Section SM-3~\cite{Note1} for details. The geometry-governed effective chiral DMI results in the domain wall phase selectivity, which is defined by signs of the topological charge $p$ and the curvature $\varkappa$. While the domain wall in the stripe with constant cross-section is pinned at the curvature maxima, the pinning position in general case is defined by complex combination \eqref{eq:q0}, see Fig.~\ref{fig:equilibrium}. In the following, for the sake of simplicity, we will consider stripes with constant aspect ratio, i.e. stripes with constant anisotropy coefficients.

Spatially varying curvature and cross-section become sources of the domain wall dynamics. Let us consider internal-to-the-system eigenmotion of the domain wall. By introducing small harmonic decaying oscillations from the equilibrium positions, we derive domain wall eigenfrequency $\omega$, normalized by  $\omega_0=4\pi \gamma_0 M_{\textsc{s}}$ as $\Omega=\omega/\omega_0$:
\begin{equation} \label{eq:omega}
\begin{aligned}
&\Omega \approx \sqrt{\Omega_{\textsc{a}}\Omega_{\text{g}}}, \qquad \Omega_{\text{g}} =  \Omega_\varkappa + \Omega_{\mathcal{S}} + \Omega_{\varkappa \mathcal{S}},\\
&\Omega_{\textsc{a}} \!=\! 2\left(k_\textsc{b} - \varkappa_0^2\right)\! + \! \frac{\pi}{\Delta_0}\! \left|\varkappa_0\right|, \; \Omega_\varkappa\! =\! - \pi \Delta_0\varkappa''_0\sign\varkappa_0,\!\!\!\\
&
\Omega_{ \mathcal{S}}=2\frac{\mathcal{S}''_0}{\mathcal{S}_0},\quad
\Omega_{\varkappa \mathcal{S}} \!= \!
-\pi\Delta_0 \frac{\sign\varkappa_0}{\mathcal{S_0}} \left(2\varkappa'_0\mathcal{S}'_0 + \varkappa_0\mathcal{S}''_0\right),
\end{aligned}
\end{equation}
where $\gamma_0$ is the gyromagnetic ratio, see Supplementary Section SM-3~\cite{Note1} for details. One can see that the expression for $\Omega_{\textsc{a}}$ essentially depends on anisotropy constants. Next terms, $\Omega_\varkappa$, $\Omega_{\mathcal{S}}$, and $\Omega_{\varkappa \mathcal{S}}$, describe the influence of the curvature deformation, cross-section deformation, and their coupling onto the eigenfrequency, respectively. 


\emph{Discussion}. 
To study the functioning of varying cross-section we deliberately separated effects of curvature and torsion from those caused by cross-section. Another instructive approach is to reduce the problem of curved magnet with varying cross-section to that of a curved stripe with a fixed cross-section, but with curvature, torsion and local anisotropy modified by the varying cross-section, for details see Supplementary Section SM-4 \cite{Note1}. In this case presented system can be treated as chiral biaxial ferromagnet.

The developed theory of curved wires and stripes allows to generalize existed theories \cite{Sheka22,Makarov22a} and to predict new effects in statics and dynamics of magnetization textures depending on the deformation of the sample cross-section. Spatially varying sample cross-section $\mathcal{S} = \mathcal{S}(\xi)$ becomes an additional source of geometry-governed DMI and anisotropy in curved wires and stripes. Even in simplest cases of rings ($\varkappa'=0$, $\sigma=0$) and helices ($\varkappa'=0$, $\sigma'=0$), the spatial deformation of the cross-section produces a coordinate-dependent Dzyaloshinskii parameter, see Supplementary Section SM-3~\cite{Note1}, which acts similar to the functionally-graded DMI \cite{Yershov20b}. One can expect similar consequences, in particular, the concept of domain wall diode \cite{Yershov20b}.

Appearance of another geometrical source of DMI is prospective for the geometrical tailoring of the magnetochirality. Curvilinear magnetism proposes the concept of mesoscale DMI which unite intrinsic DMI, determined by materials parameters and extrinsic, determined by local curvatures and torsion; both DMI influence magnetic textures acting at different lengthscales \cite{Volkov18}. Here we report on a new geometrical source of DMI, determined by the varying cross-section. The strength of this DMI contribution can be changed by properly choosing deformation of the cross-section. Such method to control DMI can find different applications, in particular to artificial magnetoelectric materials \cite{Volkov19b}.

Originated from the geometrical DMI, the varying cross-section of the sample shows itself in emergent geometry-induced magnetic field \eqref{eq:emergent-field}, and causes the tilting \eqref{eq:tilting} of equilibrium texture proportional to the cross-section gradient. Due to the chiral nature of DMI it can provide the chiral response for originally achiral system similar to the influence of the gradient of the curvature~\cite{Sheka22}.

In curvilinear magnonics, spin waves are known to be bound by the curvature gradient \cite{Gaididei18a} in curved wires. Similarly, we expect appearance of effect of localization of magnon modes by the gradient of the cross-section.

In curvilinear spintronics, the motion of domain walls in curved waveguide is essentially affected by sample curvature and torsion. In particular, the curvature gradient becomes the source of external force, which can pin the domain wall at the curvature maxima, it causes the domain wall automotion, it essentially influence domain wall mobility under external driving, for review see \cite{Sheka22,Makarov22a}. We expect series of similar effects caused by the gradient of the cross-section in 2D and 3D wires and stripes in different setups~\cite{Chauleau10,Nikonov14,Richter16,Mawass17}, including domain wall oscillations pinned by geometrical defects \cite{Alejos07}, current-induced spin-wave emitter based on pinned domain wall~\cite{Voto17}, and the domain wall automotion in curved stripes, recently observed in \cite{Skoric22}. 

\begin{figure}
	\includegraphics[width=\columnwidth]{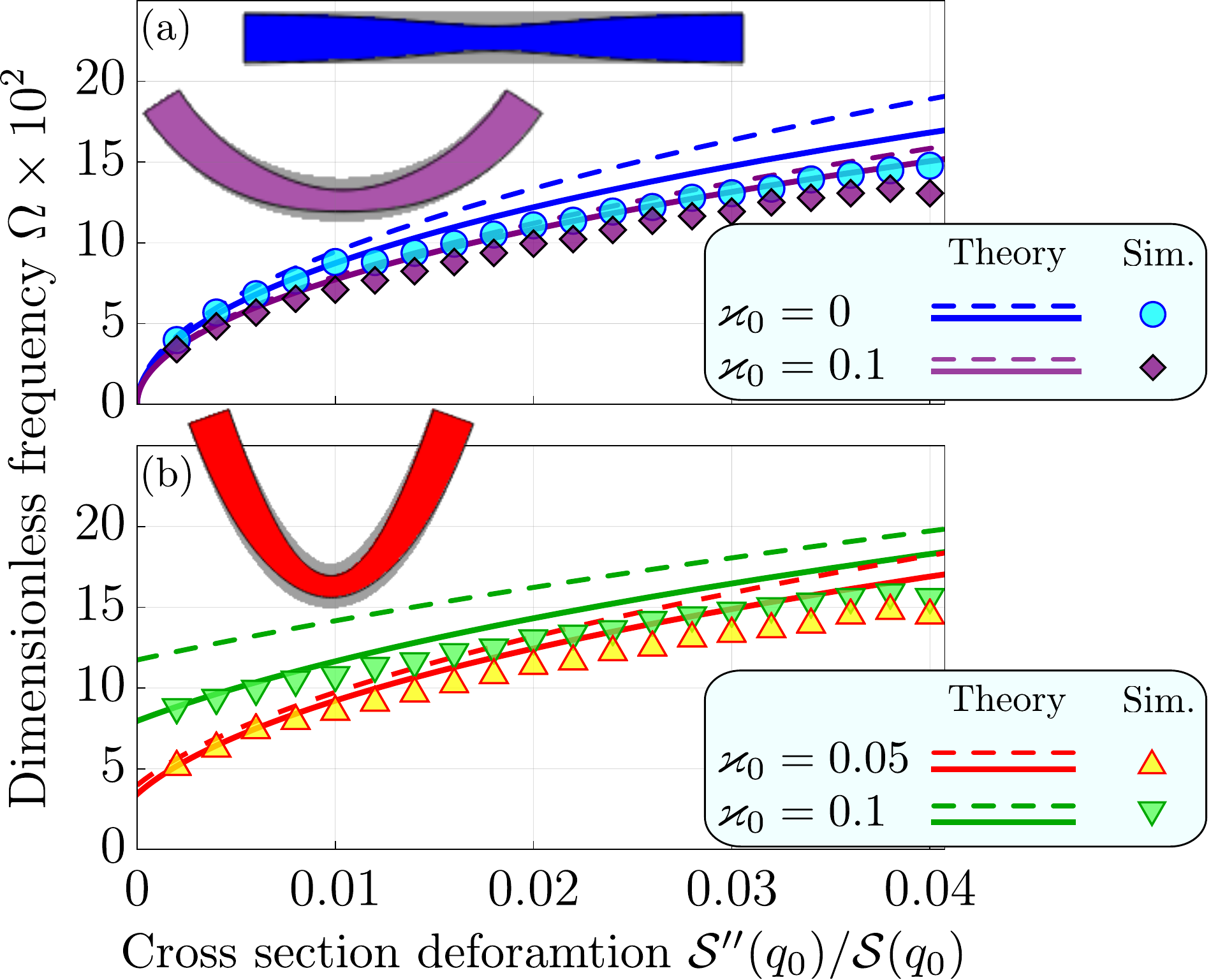}
	\caption{(Color online) \textbf{Domain wall eigenoscillations}. Frequencies for different geometries: (a) for rectilinear ($\varkappa_0=0$) and circular-segment ($\varkappa_0=0.1$) shaped stripes; (b) for parabola shaped stripes with different curvature at the extremum. Dashed lines correspond to the predictions~\eqref{eq:omega} and solid lines correspond to prediction~(S22). Symbols show the results of micromagnetic simulations, for details see Supplementary Section SM-5~\cite{Note1}.
		\label{fig:frequency}%
	}
\end{figure}

We apply our theory to the problem of domain wall statics and dynamics. First we show that the effect of domain wall pinning in planar stripe and its eigenoscillations. One can see in Fig.~\ref{fig:equilibrium} the domain wall can be pinned in geometrically unexpected place, defined by the combination between the curvature gradient and cross-section gradient, see \eqref{eq:equilibrium}. By applying weak external field or spin current, one can excite oscillation of domain wall near its equilibrium position, which is described by Eq.~\eqref{eq:omega}. Now we illustrate eigenoscillations of the domain wall by several examples. We consider three different geometries with varying cross-section: rectilinear stripe,  ring segment with constant curvature, and parabola geometry. In the case of the rectilinear stripe, the general expression for the eigenfrequency \eqref{eq:omega} is reduced to $\Omega_{\text{r}}=2 \sqrt{k_\textsc{b}\mathcal{S}''_0/\mathcal{S}_0}$ and caused by the cross-section deformation $\Omega_{ \mathcal{S}}$, see Fig.~\ref{fig:frequency}(a). In the limit case of the curved wire with a constant and circular cross-section, the well-known result $\Omega = \pi \sqrt{-\varkappa_0 \varkappa''_0}$ is reproduced \cite{Yershov15b}. The coupling between the curvature and the cross-section deformation, described by $\Omega_{\varkappa \mathcal{S}}$ stands out sharply for stripes with constant curvatures, in particular, for segments of circle, which results in $\Omega=\sqrt{\Omega_{\textsc{a}} \left(2-\pi\Delta_0|\varkappa_0|\right)\mathcal{S}''_0/\mathcal{S}_0}$, see Fig.~\ref{fig:frequency}(b). In case of small curvatures it results in $\Omega\approx \Omega_{\text{r}} \left[1-(\pi/4)|\varkappa_0|\Delta_0\left(1-{k_\textsc{t}}/{k_\textsc{b}}\right)\right]$. One can see that curvature decreases the frequency for cases with $k_\textsc{t}<k_\textsc{b}$. Temporal evolution of such domain wall oscillations in curved wire with varying cross-section is presented in a Supplementary video~\cite{Note1}. The good agreement of analytical predictions and results of full scale micromagnetic simulations for magnetically soft stripes ($K_{\textsc{t}} = 0$ and $K_{\textsc{b}}=0$) demonstrates that the approximation of magnetostatic interaction by the effective biaxial anisotropy  for thin and narrow stripes with coordinate dependent cross-section is physically sound for a domain wall dynamics.

By analyzing the properties of eigenfrequencies \eqref{eq:omega} we can make some general remarks: (i) Easy-surface anisotropy $k_\textsc{b}$ \textit{increases} the eigenfrequency of the domain wall oscillations. (ii)~Localized cross section deformation also \textit{increases} the eigenfrequency of the domain wall oscillations in the vicinity of equilibrium. (iii) However, the coupling between curvature and cross section deformation \textit{decreases} the eigenfrequencies. The corresponding conclusions are well presented in Fig.~\ref{fig:frequency}. 

To conclude, we develop a micromagnetic framework of curved wires and stripes with varying cross-section. Using this framework we proved the spatially varying cross-section becomes a new source of geometry-governed DMI on a par with local curvature and torsion. It allows us to describe effects of domain wall pinning and eigenoscillation. We expect that this theory will push new directions in curvilinear magnonics and spintronics. Its generalization for the case of curved films and shells would affect curvilinear skyrmionics as well.


\emph{Acknowledgments}. We thank Ulrike Nitzsche for technical support. We also appreciate discussions with Amalio Fernández-Pacheco and Jeroen~van den Brink. D.Sh. acknowledges the financial support by the program ``Magnetism in Ukraine Initiative'' (IEEE Magnetics Society and the Science and Technology Center of Ukraine, project number 9918). K.Y. acknowledges the financial support of the German Research Foundation (DFG) Grant No. YE 232/1-1.

%



\section*{Supplementary material}

# Supplemental Material for
# "Control of magnetic response in curved stripes by tailoring cross-section"

Kostiantyn V. Yershov[1, 2, *] and Denis D. Sheka[3, †]

[1]*Leibniz-Institut für Festkörper- und Werkstoffforschung, IFW Dresden, 01171 Dresden, Germany*
[2]*Bogolyubov Institute for Theoretical Physics of the National Academy of Sciences of Ukraine, 03143 Kyiv, Ukraine*
[3]*Taras Shevchenko National University of Kyiv, 01601 Kyiv, Ukraine*

The supplementary information provides details on analytical calculations of main aspects of the magnetization dynamics in stripes with varying cross-section.

## SM-1. MODEL

We model ferromagnet wire/stripe, which central curve $\boldsymbol{\gamma} = \{x, y, z\}$ can be parameterized by the arc length $s$. Following the notations of the main text we introduce dimensionless coordinate along this curve $\xi = s/\ell$ with $\ell = \sqrt{A/(4\pi M_s^2)}$ being the exchange length. The local reference frame at $\boldsymbol{\gamma}(\xi)$ can be introduced using a Frenet trihedron $\{\boldsymbol{e}_\text{T}, \boldsymbol{e}_\text{N}, \boldsymbol{e}_\text{B}\}$, which defines tangential, normal and binormal directions [1]. Differential properties of $\boldsymbol{\gamma}$ are described by Frenet–Serret formula

$$\boldsymbol{e}'_i = \boldsymbol{\varpi} \times \boldsymbol{e}_i, \qquad i = \text{T}, \text{N}, \text{B}, \quad (\text{S1})$$

where the Darboux vector $\boldsymbol{\varpi} = \sigma \boldsymbol{e}_\text{T} + \varkappa \boldsymbol{e}_\text{B}$ with $\varkappa = \kappa\ell$ and $\sigma = \tau\ell$ being the reduced curvature and torsion of the curve, respectively [2] ($\kappa$ and $\tau$ are dimensional curvature and torsion).

Let us defines a sample as a space domain

$$\boldsymbol{r}(\xi, \zeta_1, \zeta_2) = \boldsymbol{\gamma}(\xi) + \zeta_1 \boldsymbol{e}_\text{N}(s) + \zeta_2 \boldsymbol{e}_\text{B}(s). \quad (\text{S2})$$

We consider both wires and stripes. In case of curved stripe, the perpendicular cross-section is parameterized by $\zeta_1 \in [-w(\xi)/2, w(\xi)/2]$ and $\zeta_2 \in [-h(\xi)/2, h(\xi)/2]$ with $h(\xi)$ and $w(\xi)$ being the coordinate-dependent stripe height and width; the cross-section area $S(\xi) = w(\xi)h(\xi)$. Note that the deformation does not break an assumption of one-dimensionality of magnetization. Besides, to avoid an overlap between the neighboring windings of the stripe and reduce the influence of the demagnetization field, the distance between them must be bigger than the stripe width $w$.

We describe magnetic properties of the stripe using the model of biaxial magnet with the energy density $W = W^\text{X} + W^\text{A}$, see the main text. Here we discuss the biaxial anisotropy, $W^\text{A} = -K^\text{eff}_\text{T}(\boldsymbol{m} \cdot \boldsymbol{e}_\text{T})^2 + K^\text{eff}_\text{B}(\boldsymbol{m} \cdot \boldsymbol{e}_\text{B})^2$. Constants of effective anisotropies, $K^\text{eff}_\text{T} = K_\text{T} + 4\pi M_s^2 k^\text{ms}_\text{T}$ and $K^\text{eff}_\text{B} = K_\text{B} + 4\pi M_s^2 k^\text{ms}_\text{B}$ incorporate magnetocrystalline anisotropy coefficients $K_\text{T}$ and $K_\text{B}$ as well as magnetostatic contributions, $k^\text{ms}_\text{T}$ and $k^\text{ms}_\text{B}$. In case of straight and uniformly magnetized stripe with rectangular cross-section, the magnetostatic energy is well-known [3–5] to be reduced to effective shape anisotropy [5] with coefficients

$$\begin{aligned} k^\text{ms}_\text{T} &= \frac{\frac{1-a^2}{2a}\ln(1+a^2) + a\ln a + 2\arctan\frac{1}{a}}{2\pi}, \\ k^\text{ms}_\text{B} &= \frac{1}{2} - 2k^\text{ms}_\text{T}, \quad a = w/h \geq 1. \end{aligned} \quad (\text{S3})$$

For thin, narrow, and curved stripes the approximation of the shape anisotropy is used also for inhomogeneous magnetization states [6], including domain walls [7, 8]. In the limit case of square ($w/h = 1$) or circular cross-sections, the magnetostatic-shape-induced anisotropy coefficients (S3) are simplified to $k^\text{ms}_\text{T} = 1/4$ and $k^\text{ms}_\text{B} = 0$, which is a well known result [3, 9] including the case of curvilinear wires [10]. In case of the curved stripe with varying cross-section, its aspect ratio becomes coordinate dependent, $a = a(\xi)$, hence anisotropy coefficients becomes, in general, also coordinate dependent, $K^\text{eff}_\text{T} = K^\text{eff}_\text{T}(\xi)$ and $K^\text{eff}_\text{B} = K^\text{eff}_\text{B}(\xi)$.

In the currents study we limit ourselves by planar stripes (i.e. stripes with zero torsion). Torsion induced effects in stripes with coordinate-dependent cross section and curvature will be considered somewhere else.

## SM-2. MAGNETIZATION TILTING BY THE GRADIENT OF CROSS-SECTION

Here we discuss the case of strong anisotropy. We suppose that the magnetic texture will not deviate significantly from the equilibrium state given by the anisotropy. The geometry–induced anisotropy and DMI act as a source of the emergent geometry–induced magnetic field [11, 12], which influences even the equilibrium state magnetization: an assumed ground state becomes a subject of further modifications due to curvilinear effects [11–14]. Assuming small deviations from the tangential directions, we utilize the angular parametrization

$$\boldsymbol{m}_\text{BT} = \boldsymbol{e}_\text{T}\sin\Theta\cos\Phi + \boldsymbol{e}_\text{N}\sin\Theta\sin\Phi + \boldsymbol{e}_\text{B}\cos\Theta, \quad (\text{S4})$$

where $\Theta$ characterizes the deviation from the binormal direction and $\Phi$ describes the deviation from the tangential direction in the osculating TN plane. Then the

---
* k.yershov@ifw-dresden.de
† sheka@knu.ua

energy density (1) reads

$$\begin{aligned}\mathscr{E} = \mathcal{S}(\xi)\Big\{ &(\Theta' - \sigma \sin \Phi)^2 \\ &+ \big[\sin \Theta (\Phi' + \varkappa) - \sigma \cos \Theta \cos \Phi\big]^2 \\ &- k_{\text{T}} \sin^2 \Theta \cos^2 \Phi + k_{\text{B}} \cos^2 \Theta \Big\}.\end{aligned} \quad (S5)$$

The strictly tangential magnetisation corresponds to $\Theta_{\text{T}} = \pi/2$ and $\cos \Phi_{\text{T}} = \mathfrak{C} = \pm 1$. Assuming small deviations from this direction and putting $\Theta = \Theta_{\text{T}} + \vartheta$, $\phi = \Phi_{\text{T}} + \varphi$ with $|\vartheta|, |\varphi| \ll 1$, we can present the energy density (S5) as presented in Eq. (2).

## SM-3. DOMAIN WALL FOR PLANAR GEOMETRIES

Magnetization dynamics of this system is studied by means of phenomenological Landau–Lifshitz–Gilbert equations

$$-\sin\theta \dot\theta = \frac{\delta \mathscr{E}}{\delta \phi} + \alpha \sin^2 \theta \dot\phi, \quad \sin\theta \dot\phi = \frac{\delta \mathscr{E}}{\delta \theta} + \alpha \dot\theta, \quad (S6)$$

where constant $\alpha$ is a Gilbert damping coefficient, overdot indicates the the derivative with respect to dimensionless time $\bar{t} = \omega_0 t$, where $\omega_0 = 4\pi \gamma_0 M_{\text{s}}$ as $\Omega = \omega/\omega_0$. These equations of motion can be derived from the Lagrangian

$$\mathscr{L} = \mathscr{G} - \mathscr{E}, \quad \mathscr{G} = -\int_{-\infty}^{+\infty} \phi \sin\theta \dot\theta \mathcal{S} \, d\xi \quad (S7)$$

and Rayleigh dissipative function

$$\mathscr{R} = \frac{\alpha}{2} \int_{-\infty}^{+\infty} \big[\dot\theta^2 + \sin^2\theta \dot\phi^2\big] \mathcal{S} \, d\xi \quad (S8)$$

as Lagrange–Rayleigh equations

$$\frac{\delta \mathscr{L}}{\delta \psi_\nu} - \partial_{\bar t} \frac{\delta \mathscr{L}}{\delta \dot\psi_\nu} = \frac{\delta \mathscr{R}}{\delta \dot\psi_\nu}, \quad \psi_\nu \in \{\theta, \phi\}. \quad (S9)$$

Let us limit our consideration by planar geometries, when the torsion is absent, $\sigma = 0$. The central line of this stripe can be described by a planar curve $\boldsymbol{\gamma}(\xi) = \gamma_x(\xi)\hat{\boldsymbol{x}} + \gamma_y(\xi)\hat{\boldsymbol{y}}$. It is convenient to measure the polar angle from the anisotropy axis using

$$\boldsymbol{m}_{\text{TN}} = \boldsymbol{e}_{\text{T}} \cos\theta + \boldsymbol{e}_{\text{N}} \sin\theta \cos\phi + \boldsymbol{e}_{\text{B}} \sin\theta \sin\phi. \quad (S10)$$

Then the normalized energy of the magnetization texture in the planar geometry reads

$$\begin{aligned}\mathscr{E}[\theta, \phi] = \int_{-\infty}^\infty d\xi \mathcal{S} \Big[ &(\theta' + \varkappa \cos\phi)^2 + (\phi' \sin\theta \\ &- \varkappa \cos\theta \sin\phi)^2 + \sin^2\theta \left(k_{\text{T}} + k_{\text{B}} \sin^2\phi\right) \Big].\end{aligned} \quad (S11)$$

Equilibrium magnetization textures can be found by minimization of this energy, which results in following set of equations

$$\begin{aligned}&\theta'' - \sin\theta \cos\theta \left[\phi'^2 + \left(k_{\text{B}} - \varkappa^2\right)\sin^2\phi + k_{\text{T}}\right] \\ &- 2\phi' \varkappa \sin^2\theta \sin\phi + \varkappa \cos\phi \left(\frac{\mathcal{S}'}{\mathcal{S}} + \frac{\varkappa'}{\varkappa}\right) + \theta' \frac{\mathcal{S}'}{\mathcal{S}} = 0,\end{aligned} \quad (S12a)$$

$$\begin{aligned}&\phi'' + 2\theta' \left[\phi' \cot\theta + \varkappa \sin\phi\right] + \phi' \frac{\mathcal{S}'}{\mathcal{S}} \\ &- \left(k_{\text{B}} - \varkappa^2\right) \sin\phi \cos\phi - \varkappa \cot\theta \sin\phi \left(\frac{\mathcal{S}'}{\mathcal{S}} + \frac{\varkappa'}{\varkappa}\right) = 0.\end{aligned} \quad (S12b)$$

Eq. (S12b) has solutions $\cos\phi_0 = \mathfrak{C} = \pm 1$. Substitution of this solution into first equation in (S12) results in Eq. (4), which determines the function $\theta(\xi)$.

Now we proceed to collective variable approach, using the Ansatz (5)

$$\cos\theta^{\text{DW}} = -p \tanh\left[\frac{\xi - q(\bar t)}{\Delta}\right], \quad \phi^{\text{DW}} = \Phi(\bar t). \quad (S5')$$

Here $q(\bar t)$ and $\Phi(\bar t)$ are collective variables; the DW width $\Delta$ is assumed to be a slaved variable [3] i.e., $\Delta = \Delta[q, \Phi]$. First, we derive an effective energy of the domain wall. By substituting an Ansatz (5) into the energy functional (S11), we get the effective energy of the domain wall $\mathscr{E}^{\text{DW}} = \mathscr{E}[\theta^{\text{DW}}, \phi^{\text{DW}}]$, which results in

$$\begin{aligned}\mathscr{E}^{\text{DW}} = 2\mathcal{S}(q) \Big\{ &\frac{\hbar_1}{\Delta} + \hbar_4 k_{\text{T}}(q)\Delta + \Delta \left[\hbar_5 k_{\text{B}}(q)\right. \\ &\left. - \hbar_2 \varkappa^2(q)\right] \sin^2\Phi + p\pi \hbar_3 \varkappa(q) \cos\Phi \Big\}.\end{aligned} \quad (S13)$$

Here and below we use parameters $\hbar_i \equiv \hbar_i(q)$ defined as

$$\begin{aligned}\hbar_1 &= \frac{1}{2}\int_{-\infty}^\infty \frac{\mathcal{S}(q + x\Delta)}{\mathcal{S}(q) \cosh^2 x} dx, \\ \hbar_2 &= \frac{1}{2}\int_{-\infty}^\infty \frac{\varkappa^2(q + x\Delta)\mathcal{S}(q + x\Delta)}{\varkappa^2(q)\mathcal{S}(q)\cosh^2 x} dx, \\ \hbar_3 &= \frac{1}{\pi}\int_{-\infty}^\infty \frac{\varkappa(q + x\Delta)\mathcal{S}(q + x\Delta)}{\varkappa(q)\mathcal{S}(q)\cosh x} dx, \\ \hbar_4 &= \frac{1}{2}\int_{-\infty}^\infty \frac{k_{\text{T}}(q + x\Delta)\mathcal{S}(q + x\Delta)}{k_{\text{T}}(q)\mathcal{S}(q)\cosh^2 x} dx, \\ \hbar_5 &= \frac{1}{2}\int_{-\infty}^\infty \frac{k_{\text{B}}(q + x\Delta)\mathcal{S}(q + x\Delta)}{k_{\text{B}}(q)\mathcal{S}(q)\cosh^2 x} dx.\end{aligned} \quad (S14)$$

Equilibrium values of the domain wall position $q_0$, its phase $\Phi_0$, and the domain wall width $\Delta_0$ can be found by the energy minimization. The domain wall width $\Delta_0 = \sqrt{\hbar_{1,0}/(\hbar_{4,0} k_{\text{T},0})}$, the phase $\cos\Phi_0 = -p\,\text{sign}\,\varkappa_0$.

The corresponding value of the equilibrium domain wall position $q_0$ is determined by the equation

$$\frac{\varkappa'_0}{\varkappa_0} + \frac{S'_0}{S_0} + \frac{\hbar'_{3,0}}{\hbar_{3,0}} = \frac{\hbar_{1,0}}{\pi \hbar_{3,0} \Delta_0 |\varkappa_0|} \left( \frac{\hbar'_{1,0}}{\hbar_{1,0}} + \frac{\hbar'_{4,0}}{\hbar_{4,0}} + \frac{k'_{\text{T},0}}{k_{\text{T},0}} + 2\frac{S'_0}{S_0} \right), \quad \text{(S15)}$$

where $f_0 \equiv f(q_0)$, $f'_0 \equiv \partial_q f|_{q=q_0}$, $f_{i,0} \equiv f_i(q_0)$, and $f'_{i,0} \equiv \partial_q f_i|_{q=q_0}$. For the sake of simplicity in the following we will consider stripes with constant aspect ratio $a = \text{const}$ which results in the constant anisotropy coefficients $k_\text{T}$ and $k_\text{B}$ and, besides, equal parameters $\hbar_1 = \hbar_4 = \hbar_5$.

In the same manner we construct the effective Lagrangian of the domain wall $\mathscr{L}^{\text{DW}} = \mathscr{L}\left[\theta^{\text{DW}}, \phi^{\text{DW}}\right]$ and effective dissipative function $\mathscr{R}^{\text{DW}} = \mathscr{R}\left[\theta^{\text{DW}}, \phi^{\text{DW}}\right]$:

$$\begin{aligned} \mathscr{L}^{\text{DW}} &= 2p\hbar_1 S(q)\Phi\dot{q} - \mathscr{E}^{\text{DW}}, \\ \mathscr{R}^{\text{DW}} &= \frac{\alpha \hbar_1}{\Delta} S(q) \left( \dot{q}^2 + \Delta^2 \dot{\Phi}^2 \right). \end{aligned} \quad \text{(S16)}$$

Then effective equations of motion of the domain wall, i.e. the temporal evolution of collective coordinates $q(\bar{t})$ and $\Phi(\bar{t})$ can be described by effective Lagrange–Rayleigh equations

$$\frac{\partial \mathscr{L}^{\text{DW}}}{\partial X_\nu} - \frac{\mathrm{d}}{\mathrm{d}\bar{t}} \frac{\partial \mathscr{L}^{\text{DW}}}{\partial \dot{X}_\nu} = \frac{\partial \mathscr{R}^{\text{DW}}}{\partial \dot{X}_\nu}, \qquad X_\nu \in \{q, \Phi\}. \quad \text{(S17)}$$

Let us consider the case of narrow domain wall, $\Delta \ll \lambda_\varkappa, \lambda_S$ with $\lambda_\varkappa = \varkappa/\varkappa'$ and $\lambda_S = S/S'$ are inhomogeneity parameters for curvature and cross section deformations, respectively. In this limit case all parameters $\hbar_i = 1$, their derivatives $\hbar'_i = 0$, hence the energy of the domain wall is reduced to

$$\mathscr{E}^{\text{DW}} \approx 2S(q) \left\{ \frac{1}{\Delta} + k_\text{T} \Delta + p\pi \varkappa(q) \cos \Phi + \left[ k_\text{B} - \varkappa^2(q) \right] \Delta \sin^2 \Phi \right\}. \quad \text{(S18)}$$

In this case the general expression for the domain wall position (S15) is reduced to (6a). The effective Lagrangian and the dissipative function of the narrow domain wall

$$\begin{aligned} \mathscr{L}^{\text{DW}} &\approx 2S(q) \left\{ p\Phi\dot{q} - \frac{1}{\Delta} - k_\text{T}\Delta - p\pi\varkappa(q)\cos\Phi \right. \\ &\quad \left. - \left[ k_\text{B} - \varkappa^2(q) \right] \Delta \sin^2 \Phi \right\}, \\ \mathscr{R}^{\text{DW}} &\approx \frac{\alpha}{\Delta} S(q) \left( \dot{q}^2 + \Delta^2 \dot{\Phi}^2 \right). \end{aligned} \quad \text{(S19)}$$

By substituting the effective Lagrangian and the dissipative function into the Lagrange–Rayleigh equations (S17), one gets effective equations of motion of the narrow wall:

$$\begin{aligned} \alpha \frac{\dot{q}}{\Delta_0} + p\dot{\Phi} &= \left[ 2\varkappa(q)\Delta_0 \sin^2 \Phi - p\pi \cos \Phi \right] \varkappa'(q) \\ &\quad - \left[ \frac{2}{\Delta_0} + p\pi\varkappa(q) \cos \Phi \right] \frac{S'(q)}{S(q)}, \\ p\frac{\dot{q}}{\Delta_0} - \alpha \dot{\Phi} &= \left[ k_\text{B} - \varkappa^2(q) \right] \sin 2\Phi - p\pi \frac{\varkappa(q)}{\Delta_0} \sin \Phi. \end{aligned} \quad \text{(S20)}$$

Note that the right-hand-side in the first equation determines the internal-to-system driving force. The first term, proportional to the curvature gradient $\varkappa'(q)$ was studied previously. Namely this term is responsible for the domain wall autooscillations [15], its automotion [8]. The newcomer geometrical source is the cross-section gradient, $S'(q)$. Even in straight stripes with $\varkappa = 0$, the gradient of cross-section still results in a driving force for the domain wall.

Here, we study linear dynamics of the domain wall in vicinity of the equilibrium position. With this purpose we introduce small deviations in the way $q(\bar{t}) = q_0 + q(\bar{t})$ and $\Phi(\bar{t}) = \Phi_0 + \varphi(\bar{t})$. For the limit case of small curvature and cross section deformation the equations of motion (S20) linearized with respect to the deviations read

$$\begin{aligned} \begin{bmatrix} \dot{q} \\ \dot{\varphi} \end{bmatrix} &\approx - \begin{bmatrix} \alpha \Omega_\text{g} & -p\Delta_0 \Omega_\text{A} \\ p\Omega_\text{g}/\Delta_0 & \alpha \Omega_\text{A} \end{bmatrix} \cdot \begin{bmatrix} q \\ \varphi \end{bmatrix}, \\ \Omega_\text{g} &= \Omega_\varkappa + \Omega_S + \Omega_{\varkappa S}, \\ \Omega_\text{A} &= 2\left( k_\text{B} - \varkappa_0^2 \right) + \frac{\pi}{\Delta_0} |\varkappa_0|, \quad \Omega_S = 2\frac{S''_0}{S_0}, \\ \Omega_\varkappa &= -\pi \Delta_0 \varkappa''_0 \operatorname{sign} \varkappa_0, \\ \Omega_{\varkappa S} &= -\pi \Delta_0 \frac{\operatorname{sign} \varkappa_0}{S_0} \left( 2\varkappa'_0 S'_0 + \varkappa_0 S''_0 \right). \end{aligned} \quad \text{(S21)}$$

Set of Eqs. (S21) results in a decaying oscillations $q(\bar{t}) = q_0 \sin \Omega \bar{t}\, e^{-\eta \bar{t}}$ and $\varphi(\bar{t}) = \varphi_0 \cos \Omega \bar{t}\, e^{-\eta \bar{t}}$ with frequency $\Omega$ and friction $\eta$ defined as:

$$\Omega = \sqrt{\Omega_\text{A} \Omega_g}, \quad \eta = \frac{\alpha}{2}\left( \Omega_\text{A} + \Omega_g \right). \quad \text{(S22)}$$

In more general case $\lambda_\varkappa \Delta_0 \in (0,1)$ and $\lambda_S \Delta_0 \in (0,1)$, effective equations of the domain wall oscillations can be derived using Lagrangian and dissipative function in the form (S16) leading to the following expression for the eigenfrequency

$$\begin{aligned} \Omega &= \sqrt{\Omega_\text{A} \Omega_g}, \quad \eta = \frac{\alpha}{2}\left( \Omega_\text{A} + \Omega_g \right), \\ \Omega_\text{g} &= \Omega_\varkappa + \Omega_S + \Omega_{\varkappa S}, \\ \Omega_\text{A} &= 2\left( k_\text{B} - \frac{\hbar_{2,0}}{\hbar_{1,0}}\varkappa_0^2 \right) + \frac{\pi}{\Delta_0}\frac{\hbar_{3,0}}{\hbar_{1,0}} |\varkappa_0|, \\ \Omega_\varkappa &= -\pi \Delta_0 \frac{(\hbar_3 \varkappa)''_0}{\hbar_{1,0}} \operatorname{sign} \varkappa_0, \quad \Omega_S = 2\frac{(S\hbar_1)''_0}{S_0 \hbar_{1,0}}, \\ \Omega_{\varkappa S} &= -\pi \Delta_0 \frac{\operatorname{sign} \varkappa_0}{S_0 \hbar_{1,0}} \left[ 2\left(\hbar_3 \varkappa\right)'_0 S'_0 + \hbar_{3,0} \varkappa_0 S''_0 \right], \end{aligned} \quad \text{(S23)}$$

where $(\hbar_i \varkappa)'_0 \equiv \partial_q (\hbar_i \varkappa)|_{q=q_0}$. The corresponding eigenfrequencies of domain wall oscilations in vicinity of the equilibrium for different geometries as function of cross section deformation are presented in Fig. 3.

## SM-4. EFFECTIVE MODEL OF CURVED BIAXIAL STRIPE

In previous sections we discussed the role of varying cross-section by separating effects of curvature and torsion from effects connected with cross-section. Another approach to the same problem is to reduce the model to that of a curved stripe with a fixed cross-section, but with curvature, torsion and local anisotropy modified by the varying cross-section.

We start from the same model of curved magnet with varying cross-section (1). By applying the transformation $\zeta(\xi) = \int d\xi / \mathcal{S}(\xi)$, the total energy $\mathscr{E} = E/E_0 = \int \mathscr{E}^\star(\zeta) d\zeta$ resembles the energy of curved biaxial magnet with constant cross-section:

$$\mathscr{E}^\star = \mathscr{E}_0 + \underbrace{\mathscr{E}_A}_{\substack{\text{effective}\\\text{anisotropy}}} + \underbrace{\mathscr{E}_D}_{\substack{\text{effective}\\\text{DMI}}}, \quad \mathscr{E}_0 = \partial_\zeta m_i \partial_\zeta m_i, \quad \text{(S24)}$$
$$\mathscr{E}_A = \mathcal{K}^\star_{ij} m_i m_j, \quad \mathscr{E}_D = \varepsilon_{ijk} \mathcal{D}^\star_i m_j \partial_\zeta m_j,$$

where $\mathcal{K}^\star_{ij}(\zeta) = \mathcal{S}^2(\zeta) \mathcal{K}_{ij}$ and $\mathcal{D}^\star_i = \mathcal{S}(\zeta) \mathcal{D}_i$ are effective coordinate dependent coefficients of anisotropy and DMI, respectively. Model (S24) corresponds to a chiral biaxial ferromagnet with coordinate-dependent anisotropy coefficients and DMI.

First, we will calculate an emergent geometry-induced magnetic field (2). In new variables this field is defined as

$$\mathscr{E}^\star \approx \mathscr{E}_T - \boldsymbol{F}^\star \cdot \boldsymbol{m} + k_T^\star (\vartheta^2 + \varphi^2) + k_B^\star \vartheta, \quad \text{(S25)}$$
$$\boldsymbol{F}^\star = 2\mathfrak{C} \partial_\zeta \varkappa^\star \boldsymbol{e}_N + 2\mathfrak{C} \varkappa^\star \sigma^\star \boldsymbol{e}_B,$$

where $\varkappa^\star = \mathcal{S}(\zeta) \varkappa$, $\sigma^\star = \mathcal{S}(\zeta) \sigma$, $k_T^\star = \mathcal{S}^2(\zeta) k_T$, and $k_B^\star = \mathcal{S}^2(\zeta) k_B$ are redefined parameters of the system. As a result, the magnetization tilting is defined as

$$\vartheta \approx -\frac{\mathfrak{C}}{k_T^\star + k_B^\star} \varkappa^\star \sigma^\star, \quad \varphi \approx \frac{\partial_\zeta \varkappa^\star}{k_T^\star}. \quad \text{(S26)}$$

### A. Domain wall properties

As a next step, we consider domain wall properties within model (S24). Using curvilinear reference frame, one can parameterize the magnetization as $\boldsymbol{m} = \boldsymbol{e}_T \cos\theta + \boldsymbol{e}_N \sin\theta \cos\phi + \boldsymbol{e}_B \sin\theta \sin\phi$, where $\theta = \theta(\zeta)$ and $\phi = \phi(\zeta)$ are magnetic angles. The minimization of the energy results in a planar texture within the stripe plane with $\cos\phi = \mathfrak{C} = \pm 1$ and planar deviations from the tangential direction described by $\theta(\zeta)$, which satisfies the driven nonlinear pendulum equation

$$\partial_{\zeta\zeta}\theta - k_T^\star \cos\theta \sin\theta = f^\star(\zeta), \quad f^\star(\zeta) = -\mathfrak{C} \partial_\zeta \varkappa^\star. \quad \text{(S27)}$$

The spatially dependent external force $f^\star(\zeta)$ results in the absence of strictly tangential magnetization pattern.

In order to analyze domain wall properties we apply a collective variable approach based on the $q-\Phi$ model [16, 17]

$$\cos\theta^{\text{DW}} = -p \tanh\left[\frac{\zeta - q(t)}{\Delta}\right], \quad \phi^{\text{DW}} = \Phi(t). \quad \text{(S28)}$$

When the force is absent, $f^\star(\zeta) = 0$, this model provides an exact solution of (S27); it describes head-to-head or tail-to-tail domain walls with domain wall width $\Delta = 1/\sqrt{k_T^\star}$ for $p = 1$ and $p = -1$, respectively.

By substituting Ansatz (S28) into the energy density (S24) and performing integration over the $\zeta$ we get an effective domain wall energy (for the case of narrow domain wall)

$$\frac{\mathscr{E}_\star^{\text{DW}}}{2} \approx \frac{1}{\Delta} + k_T^\star(q)\Delta + p\pi \varkappa^\star(q) \cos\Phi \\ + \left\{ k_B^\star - [\varkappa^\star(q)]^2 \right\} \sin^2\Phi. \quad \text{(S29)}$$

Equilibrium values of the domain wall position $q_0$, its phase $\Phi_0$, and the domain wall width $\Delta_0$ can be found by the energy minimization. The corresponding values are determined by the equations

$$\frac{(\partial_\zeta k_T^\star)_0}{\sqrt{k_{T,0}^\star}} = \pi \operatorname{sign} \varkappa_0^\star (\partial_\zeta \varkappa^\star)_0, \quad \text{(S30)}$$
$$\Delta_0 = \frac{1}{\sqrt{k_{T,0}^\star}}, \quad \cos\Phi_0 = -p \operatorname{sign} \varkappa_0^\star.$$

where $(\partial_\zeta g^\star)_0 \equiv \partial_\zeta g^\star|_{q=q_0}$ and $k_{T,0}^\star = k_T^\star(q_0)$.

In the same manner we construct the effective Lagrangian of the domain wall $\mathscr{L}_\star^{\text{DW}} = \mathscr{L}[\theta^{\text{DW}}, \phi^{\text{DW}}]$ and effective dissipative function $\mathscr{R}_\star^{\text{DW}} = \mathscr{R}[\theta^{\text{DW}}, \phi^{\text{DW}}]$:

$$\mathscr{L}_\star^{\text{DW}} = 2p\Phi\dot{q} - \mathscr{E}_\star^{\text{DW}}, \\ \mathscr{R}_\star^{\text{DW}} = \frac{\alpha}{\Delta}\left(\dot{q}^2 + \Delta^2 \dot{\Phi}^2\right). \quad \text{(S31)}$$

By substituting the effective Lagrangian and the dissipative function into the Lagrange–Rayleigh equations (S17), one gets effective equations of motion of the narrow wall:

$$\alpha\frac{\dot{q}}{\Delta_0} + p\dot{\Phi} = \left[2\varkappa^\star(q)\Delta \sin^2\Phi - p\pi \cos\Phi\right]\partial_q \varkappa^\star(q) \\ - \Delta\left[\partial_q k_T^\star(q) + \sin^2\Phi \partial_q k_B^\star(q)\right], \\ p\frac{\dot{q}}{\Delta_0} - \alpha\dot{\Phi} = \left\{k_B^\star - [\varkappa^\star(q)]^2\right\}\sin 2\Phi - p\pi \frac{\varkappa^\star(q)}{\Delta_0}\sin\Phi. \quad \text{(S32)}$$

Equations of motion (S32) linearized within the equilibrium state (S30) of domain wall reads as

$$\begin{bmatrix}\dot{q}\\\dot{\varphi}\end{bmatrix} \approx -\begin{bmatrix}\alpha\Omega_g^\star & -p\Delta_0\Omega_A^\star \\ p\Omega_g^\star/\Delta_0 & \alpha\Omega_A^\star\end{bmatrix} \cdot \begin{bmatrix}q\\\varphi\end{bmatrix},$$
$$\Omega_g^\star = \frac{(k_T^\star)''_0}{k_{T,0}^\star} - \pi\Delta_0 \operatorname{sign}\varkappa_0^\star (\varkappa^\star)''_0, \quad \text{(S33)}$$
$$\Omega_A^\star = 2\left[k_B^\star - (\varkappa_0^\star)^2\right] + \frac{\pi}{\Delta_0}|\varkappa_0^\star|,$$

where $(g^\star)''_0 = \partial_{qq}(g^\star)|_{q=q_0}$.

Set of Eqs. (S33) results in a decaying oscillations $q(\bar{t}) = q_0 \sin\Omega\bar{t}\, e^{-\eta\bar{t}}$ and $\varphi(\bar{t}) = \varphi_0 \cos\Omega\bar{t}\, e^{-\eta\bar{t}}$ with frequency $\Omega$ and friction $\eta$ defined as:

$$\Omega = \sqrt{\Omega^\star_{\rm A}\Omega^\star_g}, \quad \eta = \frac{\alpha}{2}\left(\Omega^\star_{\rm A} + \Omega^\star_g\right). \quad (S34)$$

## SM-5. NUMERICAL SIMULATIONS

In order to verify our analytical results we perform numerical micromagnetic simulations of the Landau–Lifshitz–Gilbert equation utilizing the Nmag code [18]. We restrict ourselves to the case of magnetically soft material; therefore only two magnetic interactions are taken into account, namely exchange and magnetostatic contributions. In simulations we use material parameters of permalloy, namely, exchange constant $A = 2.6$ $\mu$erg/cm (in SI units $A^{\rm SI} = 26$ pJ/m), saturation magnetization $M_s = 800$ G (in SI units $M_s^{\rm SI} = 800$ kA/m ). These parameters result in the exchange length $\ell \approx 5.7$ nm and $\omega_0 = 28.15$ GHz. Thermal effects and anisotropy are neglected. An irregular tetrahedral mesh with cell size about 2.75 nm is used.

In all simulations we considered stripes with total length $L = 500$ nm and cross section defined as $\mathcal{S}(\xi) = \mathfrak{s}_0\{1 - \varrho/\cosh[(\xi - \eta)/\lambda]\}$. Parameters $\mathfrak{s}_0 = 75/\ell^2$ and aspect ratio $a = 3$ are fixed for all simulations.

We consider three different geometries: (i) rectilinear stripe with $\varkappa = \varkappa' = 0$; (ii) circle segment with curvature $\varkappa_0 = 0.1$ and $\varkappa' = 0$ (for circle geometry we did not consider closed loop in order to have only one domain wall); (iii) the parabola geometry $\boldsymbol{\gamma} = x\hat{\boldsymbol{x}} + \varkappa_0 x^2\hat{\boldsymbol{y}}/(2\ell)$ with $\varkappa_0 = 0.05$ and $\varkappa_0 = 0.1$ being the extreme curvature.

The numerical experiment consists of three steps. Initially, we relax the domain wall structure in an overdamped regime ($\alpha = 0.25$) in order to determine the equilibrium values of collective variables: position $q_0$ and phase $\Phi_0$. The obtained results fully coincide with the prediction (S15). To determine the values of $q$ and $\Phi$ we extract the curvilinear magnetization components $m_{\rm T} = \boldsymbol{m}\cdot\boldsymbol{e}_{\rm T}$, $m_{\rm N} = \boldsymbol{m}\cdot\boldsymbol{e}_{\rm N}$, and $m_{\rm B} = \boldsymbol{m}\cdot\boldsymbol{e}_{\rm B}$ from the simulation data, and apply fitting with the Ansatz (5). Namely, the position $q$ is determined as a fitting parameter for the function $m_{\rm T}(\xi) = -p\tanh[(\xi - q)/\delta]$, then the phase is determined from the equation $\tan\Phi = m_{\rm B}(q)/m_{\rm N}(q)$.

In the second step we slightly perturb the domain wall phase $\Phi$ by applying a weak magnetic field $\boldsymbol{B} = B_0\hat{\boldsymbol{z}}$ perpendicularly to the stripe plane, where $B_0 = 15$ mT. After the system relaxation in the applied field $\boldsymbol{B}$ the field is switched off and the magnetization dynamics is simulated for the natural value of the damping coefficient $\alpha = 0.01$ (the third step). Since the variables $q$ and $\Phi$ are canonically conjugated, see Eqs. (S20), the perturbed dynamics of $\Phi$ induces the dynamics of $q$, i.e., the domain wall starts to move, for details see movies:

1. `parabola_stripe_eta_0.mp4` shows domain wall oscillations in a parabola shaped stripe with $\varrho = 0.9$, $\eta = 0$, and $\lambda = 15$ and maximal curvature $\varkappa_0 = 0.2$.

2. `parabola_stripe_eta_7.5.mp4` shows domain wall oscillations in a parabola shaped stripe with $\varrho = 0.9$, $\eta = 7.5$, and $\lambda = 15$ and maximal curvature $\varkappa_0 = 0.2$.

3. `circle_segment.mp4` shows domain wall oscillations in a circle segment shaped stripe with $\varrho \approx 0.67$, $\eta = 0$, $\lambda = 10$, and curvature $\varkappa_0 = 0.05$.

4. `straight_stripe.mp4` shows domain wall oscillations in a straight stripe with $\varrho = 0.9$, $\eta = 0$, and $\lambda = 15$.

In all cases Gilbert damping parameter was fixed $\alpha = 0.01$.



\end{document}